# Negative Chemical Pressure Effect on Superconductivity and Charge Density Wave of $Cu_{0.5}Ir_{1-x}Zr_xTe_2$

*Lingyong Zeng[a], Yi Ji[b], Dongpeng Yu[a], Shu Guo[c,d], Yiyi He[a], Kuan Li[a], Yanhao Huang[a], Chao Zhang[a], Peifeng Yu[a], Shaojuan Luo[e], Huichao Wang[b], Huixia Luo[a*]*

[a]School of Materials Science and Engineering, State Key Laboratory of Optoelectronic Materials and Technologies, Key Lab of Polymer Composite & Functional Materials, Guangzhou Key Laboratory of Flexible Electronic Materials and Wearable Devices, Sun Yat-Sen University, No. 135, Xingang Xi Road, Guangzhou, 510275, P. R. China

[b]School of Physics, Sun Yat-Sen University, No. 135, Xingang Xi Road, Guangzhou, 510275, P. R. China

[c]Shenzhen Institute for Quantum Science and Engineering, Southern University of Science and Technology, Shenzhen 518055, China

[d]International Quantum Academy, Shenzhen 518048, China.

[e]School of Chemical Engineering and Light Industry, Guangdong University of Technology, Guangzhou, 510006, P. R. China

*Corresponding author/authors complete details (Telephone; E-mail:) (+86)-2039386124; E-mail address: luohx7@mail.sysu.edu.cn;

**ABSTRACT** This study demonstrates the design and synthesis of $Cu_{0.5}Ir_{1-x}Zr_xTe_2$ ($0 \leq x \leq 0.15$) system by partial substitution of Ir with Zr acting as a negative chemical pressure. With the doping of Zr, the cell parameters significantly expand, signifying an effective negative chemical pressure. The experimental results found evidence that the charge density wave (CDW)-like order is immediately quenched by subtle Zr substitution for Ir and a classical dome-shape $T_c(x)$ that peaked at 2.80 K can be observed. The optimal $Cu_{0.5}Ir_{0.95}Zr_{0.05}Te_2$ compound is a BCS-type superconductor and exhibits type-II SC. However, high Zr concentration ($x > 0.1$) can provoke disorder, inducing the reappearance of CDW order. The present study shows that the $Cu_{0.5}Ir_{1-x}Zr_xTe_2$ ($0 \leq x \leq 0.15$) system may provide a new platform for further understanding of multiple electronic orders in transition metal dichalcogenides.

**INTRODUCTION**

The strong coupling of the spin, orbit and lattice under pressure induces novel phenomena into solid-state materials, such as ferromagnetism, multiferroic, superconductivity (SC), etc[1-13]. In the category of origin, pressure can be classified into inner chemical pressure and external physical pressure. High external pressure has become a powerful means to create and control high-temperature SC. Recently, a sensational achievement with this method is the achievement of room temperature SC in the carbonaceous sulfur hydride, at the expense of high pressure of ≈ 270 GPa[14]. However, rare experimental teams can achieve such high pressure. Chemical pressure as internal pressure is another efficient method for tailoring functional materials and even inducing alluring quantum states. It is often regarded as exerted force due to the lattice strain by heterogeneous substitution in the lattice. Furthermore, the influence of negative pressure (expansion of lattice) and positive pressure (compression of lattice) may offer novel insights for the structural instabilities, which are valuable to turn the SC or other exceptional phenomena in solid materials.

In cuprates and the iron-based high-$T_c$ superconductors, SC is induced by doping charge carriers into the pristine compound to suppress the charge density wave (CDW) or antiferromagnetic state[15-20]. The same is true for the low dimensional transition metal dichalcogenides (TMDCs) superconductors where SC is observed through intercalation or substitution of the pristine CDW compounds[21-25]. Again, the $T_c$ can be modified over a wide doping range with a highest $T_c$ at the optimal doping content. Revealing this evolution also serves as a pre-requisite for uncovering the origin of high-$T_c$ in unconventional superconductors. Doping due to injection of charge carriers (holes or electrons), chemical pressure (positive or negative), and related non-magnetic point-like disorder modify the Fermi surface resulting in affect superconducting critical temperature ($T_c$) and CDW transition temperature ($T_{CDW}$).

Two-dimensional layered compound $IrTe_2$ displays a structural phase transition from a high temperature trigonal to a low temperature monoclinic phase at around 280 K [26-28]. Previous studies have shown that the phase transition can be suppressed and SC was induced by chemical doping [28-31]. Due to the weak interlayer force, Cu can be intercalated in $IrTe_2$ layers and a small amount of Cu intercalation also suppresses the phase transition and emerges SC in $Cu_xIrTe_2$[32]. Moreover, when $x = 0.5$, an anomaly in the temperature dependence of resistivity and magnetic susceptibility is observed [26]. Recently, quasi-two-dimensional $Cu_{0.5}IrTe_2$ features not only a $T_{CDW}^{Cooling} \sim 186$ K with characteristic CDW-related hump in resistance, but also a $T_c \sim 2.5$ K[33]. The energy bands near the Fermi level of $Cu_{0.5}IrTe_2$ are derived from Ir $d$ and Te $p$ orbitals and locates at a flat plateau. Recently, studies show that the substitution of Ru or Ti ions for Ir in $Cu_{0.5}IrTe_2$ suppress the CDW-like state and form a dome-shape superconducting region[34,35]. The decrease in the volume of the lattice parameters due to the substitution of Ru or Ti ions may be interpreted as a positive chemical pressure. Nevertheless, it is still a lack of study of negative chemical pressure induced by Ir-site substitution in the $Cu_{0.5}IrTe_2$ system. To introduce high negative chemical pressure inside $Cu_{0.5}IrTe_2$, a facile and straightforward approach is to substitute Ir with larger Zr atoms. Also, Zr has been used as a dopant to enhance the up critical field ($H_{c2}$) in $MgB_2$ superconductor and improve critical current density ($J_c$) in copper-based superconductors (e.g., $Gd_2Ba_4CuO_{7-\delta}$, $YBa_2Cu_3O_{7-\delta}$)[36-39]. The further theoretical calculation indicates that Zr doping can improve the superconducting properties of metal-hydride $YH_2$ in the absence of applied pressure[40]. In such circumstances, it remains the meaningful issue as to whether the SC and CDW can be turned by introducing negative chemical pressure in $Cu_{0.5}IrTe_2$ system.

Here, we report the effect of large Zr atoms substitution for Ir on the physical properties of $Cu_{0.5}Ir_{1-x}Zr_xTe_2$, where the increase of $x$ in $Cu_{0.5}Ir_{1-x}Zr_xTe_2$ 1) expands the lattice constant, hence inducing the effective negative chemical pressure ($P_{ch} \leq 0$); 2) suppresses the CDW order immediately; 3) forms a dome-shaped superconducting phase diagram vs. doping $x$; 4) drives the system toward disorder at critical Zr concentration of $x = 0.125$ and causes the reappearance of CDW order. These findings add new members for the chalcogenide superconductors and also might provide new sights into design new superconducting compounds.

**EXPERIMENT**

Polycrystalline compounds of composition $Cu_{0.5}Ir_{1-x}Zr_xTe_2$ ($0 \leq x \leq 0.15$) were synthesized by the solid-state reaction method. Before mixing the raw materials, we will purify the Cu powders at 600 °C for 6 - 12 hours under 5% $H_2$ + 95% Ar atmosphere. Then Cu (99.99%), Ir (99.99%), Zr (99.9%), and Te (99.99%) powder with the ratio of 0.5 : 1-$x$ : $x$ : 2.05 were mixed together and sealed in the evacuated quartz tubes. The samples were put in a box furnace at 850 °C for 5 days. The obtained powders were then ground, pelletized, and heated under the same condition for several times. Powder x-ray diffraction (XRD) was examined by Cu *Ka1* radiation on MiniFlex, Rigaku. Rietveld refinements were performed with the Fullprof Suite software. The electrical resistivity measurements were carried out with a standard four-contact method on a Quantum Design physical property measurement system (PPMS). The magnetic measurements and heat capacity measurement were carried out on the same instrument.

**RESULTS AND ANALYSIS**

The powder XRD patterns of $Cu_{0.5}Ir_{1-x}Zr_xTe_2$ polycrystalline samples are displayed in **Figure 1**. All the peaks are perfectly indexed with a tetragonal structure ($P\bar{3}m1$ space group), except for several tiny peaks due to the unreacted impurity Ir in low concentration Zr doped samples ($0 \leq x \leq 0.05$). This confirms that no structural phase transition happens in $Cu_{0.5}Ir_{1-x}Zr_xTe_2$ ($0 \leq x \leq 0.15$) system. The refined unit cell parameters are plotted as a function of Zr content in **Figure 1**. For the pristine $Cu_{0.5}IrTe_2$, the *a* and *c* lattice parameters are found to be 3.9397(5) Å and 5.3965(3) Å, respectively[26]. With increasing Zr content, both *a* and *c* expand, which causes the shift of primary (001) peak towards lower Bragg angle, as displayed in the top left corner inset of **Figure 1**. The increase in the lattice parameters is as expected since the atomic radius of Zr (160 pm) is larger than that of Ir (135.5 pm)[41]. Therefore, Zr doping in the $Cu_{0.5}IrTe_2$ compound is like the application of negative pressure on the system. Moreover, the maximum doping amount of Zr is 0.15 in this system since impurity $Ir_3Te_8$ impurity is detected when $x > 0.15$. **Figure 2** displays the Rietveld refinement of the representative $Cu_{0.5}Ir_{0.9625}Zr_{0.0375}Te_2$ compound. The inset in **Figure 2** shows the $Cu_{0.5}Ir_{1-x}Zr_xTe_2$ crystal structure, in which Cu intercalated between two-dimensional (2D) layers consisting of covalently bonded Ir and Te atoms, and Zr partial substitutes Ir. We further used SEM-EDXS to check the elements distribution in our samples (see **Figure S1**). The SEM-EDXS results suggest that the ratios of our obtained samples are close to the target-designed compounds (see **Table S1**).

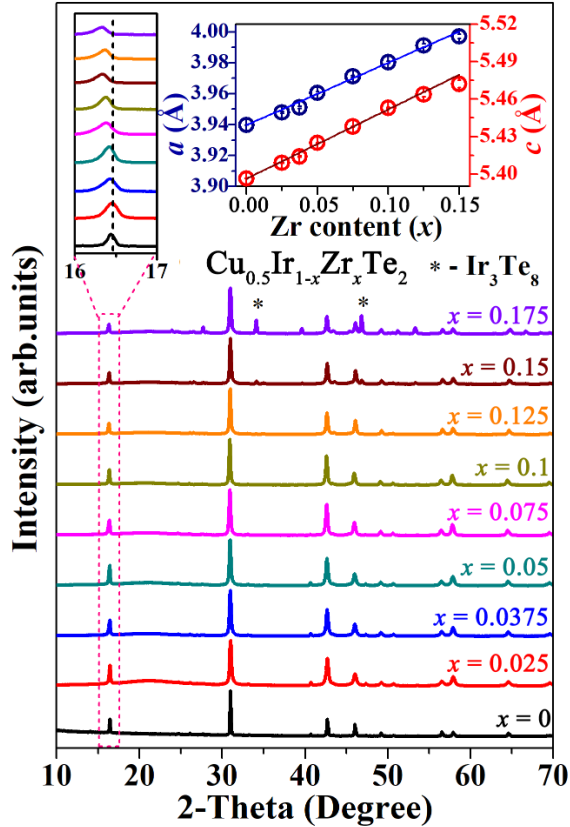

**Figure 1.** Powder XRD patterns of $Cu_{0.5}Ir_{1-x}Zr_xTe_2$ ($0 \leq x \leq 0.15$) samples. The left inset shows a shift to low angles of the (001) peak as a function of the Zr content. The right inset shows the evolution of the lattice parameters indicating an expansion of the unit cell.

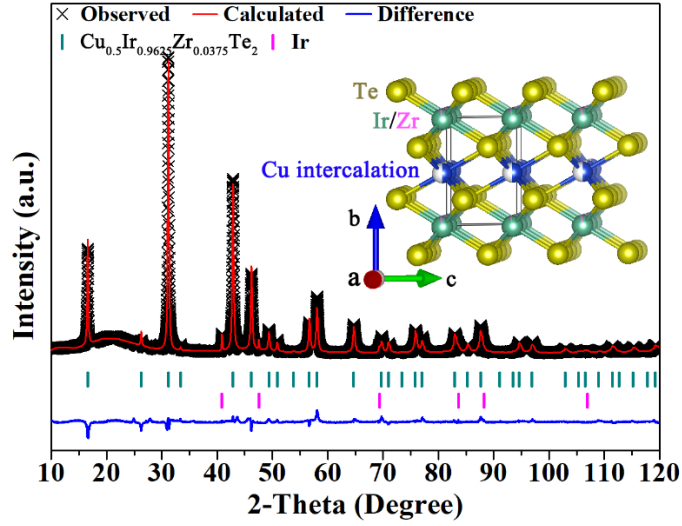

**Figure 2.** Observed and calculated XRD patterns of $Cu_{0.5}Ir_{0.9625}Zr_{0.0375}Te_2$. Inset displays the crystal structure for $Cu_{0.5}Ir_{1-x}Zr_xTe_2$ samples with space group $P\bar{3}m1$.

**Table 1**. Superconducting transition temperature ($T_c$), residual resistance ratio (RRR), and CDW transition temperature ($T_{CDW}$) of polycrystalline $Cu_{0.5}Ir_{1-x}Zr_xTe_2$ ($0 \leq x \leq 0.15$) samples.

| Zr content ($x$) | $T_c$ (K) | RRR ($R_{300 K}/R_{3 K}$) | $T_{CDW}$ (K) (cooling) |
|---|---|---|---|
| 0 | 2.42 | 4.14 | 186 |
| 0.025 | 2.65 | 5.00 | - |
| 0.0375 | 2.70 | 5.15 | - |
| 0.05 | 2.80 | 5.08 | - |
| 0.075 | 2.74 | 4.35 | - |
| 0.1 | 2.70 | 4.55 | - |
| 0.125 | 2.49 | 2.41 | 232 |
| 0.15 | 2.17 | 2.92 | 239 |

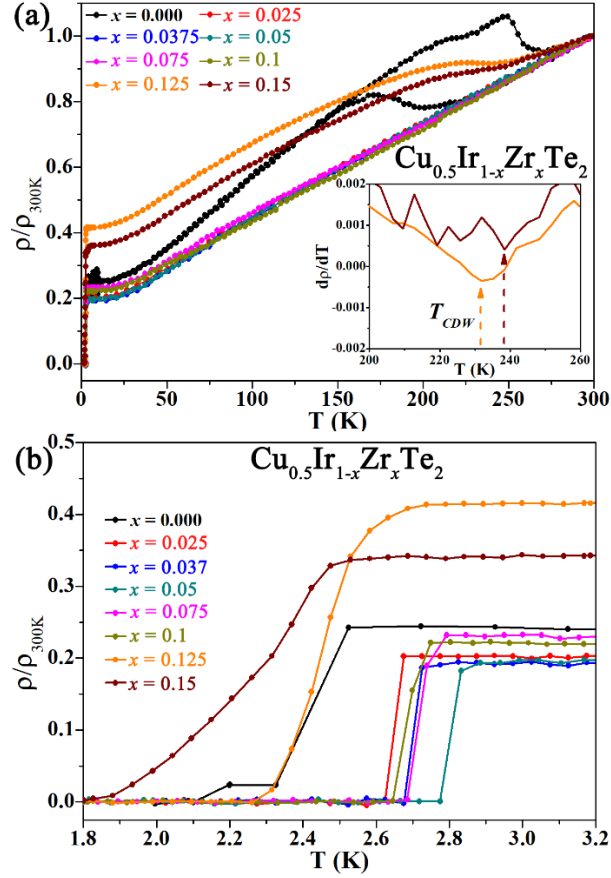

**Figure 3.** (a) Temperature dependence of normalized $\rho/\rho_{300K}$ of $Cu_{0.5}Ir_{1-x}Zr_xTe_2$ system. (b) Temperature dependence of normalized $\rho/\rho_{300K}$ of $Cu_{0.5}Ir_{1-x}Zr_xTe_2$ ($0 \leq x \leq 0.15$) samples at 1.8 - 3.2 K.

The temperature dependence of resistivity for the $Cu_{0.5}Ir_{1-x}Zr_xTe_2$ system are shown in **Figure 3a, b**. On the cooling below room temperature, the undoped sample $CuIr_2Te_4$ exhibits a metallic behavior but undergoes a CDW-like hump around 186 K due to partial opening of a gap at the Fermi surface and the consequent reduction in density of states (DOS). CDW transition temperature $T_{CDW}$ was assigned as the minimum of the temperature derivative of resistivity (d$\rho$/dT), which is displayed in the inset of **Figure 3a**. When Zr is introduced into the system, the CDW-like hump is rapidly quenched upon slight Zr doping ($x$ = 0.025). Small amount of Zr doping can suppress the CDW order and enhance the $T_c$. If we notice that the electro-negativity of Zr (1.33) is much smaller than that of Ir (2.2)[42], it can be foreseen that the substitution of Ir by Zr weakens the chemical bonding of Ir-Te, thereby might improving the itinerancy of conduction on the $IrTe_2$ layer. One of the results is that the concentration of conduction electron will increase, which suppressing the instability of CDW and the enhancement of SC. Similarly, the phenomenon of modifying CDW and SC states by weakening the chemical bonding are also observed in the $Eu_3Bi_2S_{4-x}Se_xF_4$ system[43].

**Figure 3b** shows ρ(T) below 3.2 K of all the $Cu_{0.5}Ir_{1-x}Zr_xTe_2$ samples. The $T_c$ was taken as the midpoint of the resistivity transitions. We can see that the $T_c$ increases slightly and reaches the maximal value (2.8 K) at $x$ = 0.05, accompanied by a decrease in $T_c$ for higher Zr doping content $x$. The superconducting transition width ($\Delta T_c$) is quite large in both two last doping compounds ($x$ = 0.125, 0.15) where the coexistence of CDW order and SC can be observed. Generally, the CDW will induce a band gap, thereby reducing the average DOS at the Fermi surface, which is unfavorable to SC[45]. Furthermore, the absence of CDW order samples (0 < $x$ ≤ 0.1) have a large residual resistivity ratio (RRR = $R_{300K}/R_{3K}$) than that of the pristine sample and exhibit sharper superconducting transition at $T_c$. The $T_c$ and RRR values of $Cu_{0.5}Ir_{1-x}Zr_xTe_2$ (0 ≤ $x$ ≤ 0.15) samples are summarized in **Table 1**. When $x$ ≥ 0.125, interestingly, the CDW-like hump reappears, and $T_{CDW}$ increases upon increasing $x$. The increase of Zr concentration $x$ in $Cu_{0.5}Ir_{1-x}Zr_xTe_2$ will lead to the increment of the lattice parameters in the sample, followed by the increase of the strain, grain boundaries, and other crystallographic defects. The aforementioned factors all can affect RRR. Correspondingly, the lower values of RRR = 2.41, and 2.92 for the $x$ = 0.125, and 0.15 samples imply more disorder in these two samples, which is likely to cause the reemergence of the CDW. The rapid decrease of RRR caused by negative chemical pressure is also found to promote the CDW order in 2H-NbSe$_{2-x}$Te$_x$ system[46].

We further carried out Hall measurements of three different Cu$_{0.5}$Ir$_{1-x}$Zr$_x$Te$_2$ ($x$ = 0.0325, 0.075, 0.1) samples. As shown in **Figure S2a**, the hall traces show liner dependence on the perpendicular magnetic field at different selected temperatures form 2 - 200 K. The nearly identical slope of different samples indicates that the carrier concentration does not show obvious change with increasing Zr doping. Moreover, the carrier density shown in **Figure S2b** of the Zr doped samples are comparable to that of undoped Cu$_{0.5}$IrTe$_2$[44]. Thus, the Hall results suggests that the carriers are not doped by Zr substitution in this system.

**Figure 4a** displays the temperature dependence of the resistivity of Cu$_{0.5}$Ir$_{1-x}$Zr$_x$Te$_2$ system. As a multiband electronic system with local CDW fluctuations above $T_{CDW}$, the carrier scattering mechanism arises from collective excitations blow the CDW and from local CDW fluctuations above the CDW. Above $T_{CDW}$, resistivity $\rho(T)$ follows the relation of $\rho(T) \sim AT + B$. If temperature is immediately below $T_{CDW}$, $\rho(T) \sim CT^2$, while $\rho(T) \sim DT^5$ when the temperature is below $\sim$ 20 K[46]. The T$^5$ is caused by normal electron-phonon scattering, whereas the T$^2$ arises due to the scattering of electrons by collective excitations of CDW. The linear term A and B above $T_{CDW}$ are caused by electron-phonon scattering and phase disorder impurity-like scattering due to local CDW fluctuations. **Figure 4b** displays the electronic scattering mechanism for the Cu$_{0.5}$Ir$_{0.875}$Zr$_{0.125}$Te$_2$ compound. The inset of **Figure 4b** displays the detail resistivity fitted with $\rho(T) \sim DT^5$ at the temperature range of $T_c$ to 15 K. And the term B fits of resistivity for Cu$_{0.5}$Ir$_{1-x}$Zr$_x$Te$_2$ ($0 \leq x \leq 0.15$) samples are summarized in **Figure 4c**. Obviously, the resistivity is linear at high temperatures in Cu$_{0.5}$Ir$_{1-x}$Zr$_x$Te$_2$ ($0 \leq x \leq 0.15$) samples. For comparison, the RRR value of Cu$_{0.5}$Ir$_{1-x}$Zr$_x$Te$_2$ ($0 \leq x \leq 0.15$) samples are also displayed in **Figure 4c**. It can be clearly visible that the trend of the change of RRR value and term B with the Zr content is just the opposite. Furthermore, the rapid decline in the RRR value and the rapid increase in the term B value exactly where the CDW order reappears, implying strong sensitivity of CDW to Zr substitution for Ir with a negative chemical pressure and possibly to disorder.

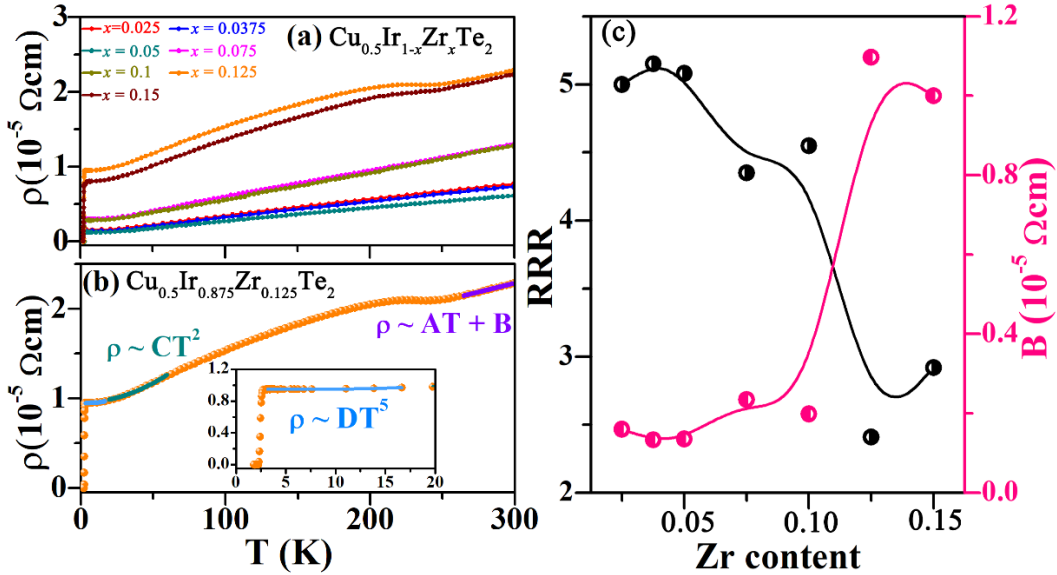

**Figure 4** (a) Temperature dependence of the resistivity for $Cu_{0.5}Ir_{1-x}Zr_xTe_2$ system. (b) Electronic scattering mechanism for $Cu_{0.5}Ir_{0.875}Zr_{0.125}Te_2$. (c) Zr content dependence of the residual resistance ratio (RRR) and the term B of fitting resistivity.

Correspondingly, the diamagnetic signals appear in magnetic susceptibility $4\pi\chi(T)$ curves at 10 Oe (see **Figure 5a**). The zero-field-cooling (ZFC) $4\pi\chi(T)$ curves display strong diamagnetic responses because of the Meisner effect, indicating bulk SC for our studied polycrystalline $Cu_{0.5}Ir_{1-x}Zr_xTe_2$ ($0 \leq x \leq 0.1$) samples. The diamagnetic response is weakened when $x > 0.1$, in good agreement with large $\Delta T_c$ in the resistivity measurement. Furthermore, to confirm the presence of a CDW-like transition in the polycrystalline $Cu_{0.5}Ir_{0.85}Zr_{0.15}Te_2$ compound, magnetic susceptibility has been carried out under the applied fields of 10 KOe as presented in **Figure 5b**. The magnetic susceptibility with cooling and heating displays a distinct hysteresis associated with the formation of CDW-like transition, which is in good agreement with the CDW hump found in resistivity measurement. The $T_{CDW} \approx 240$ K obtained from the midpoint of sudden hysteresis transition, which is also consistent with the resistivity data.

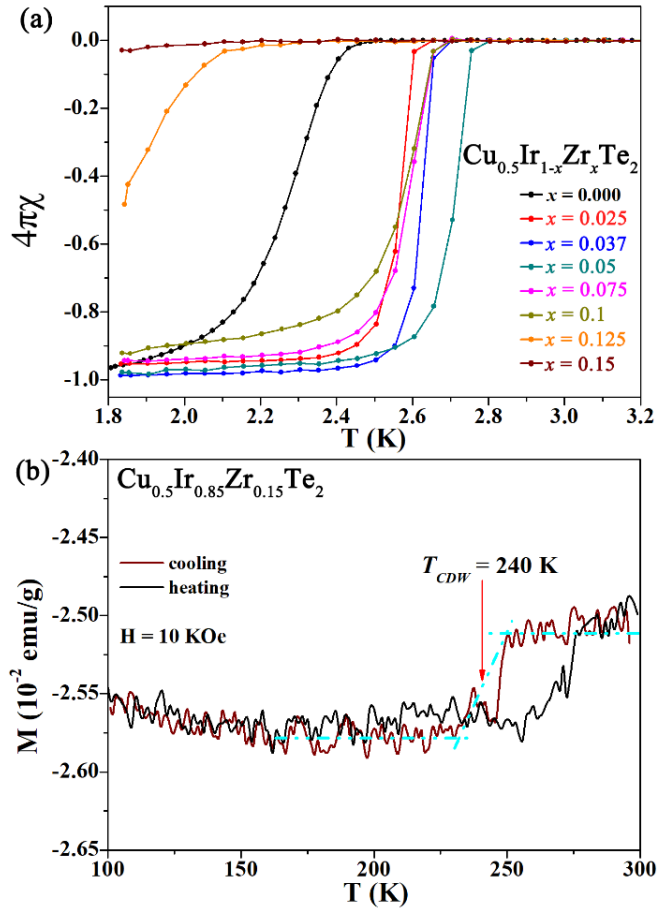

**Figure 5** Magnetic characterization for $Cu_{0.5}Ir_{1-x}Zr_xTe_2$ system. (a) The zero-field cooling magnetic susceptibilities for $Cu_{0.5}Ir_{1-x}Zr_xTe_2$ samples under the applied field 10 Oe. (b) Magnetization curves measured under 10 KOe for $Cu_{0.5}Ir_{0.85}Zr_{0.15}Te_2$.

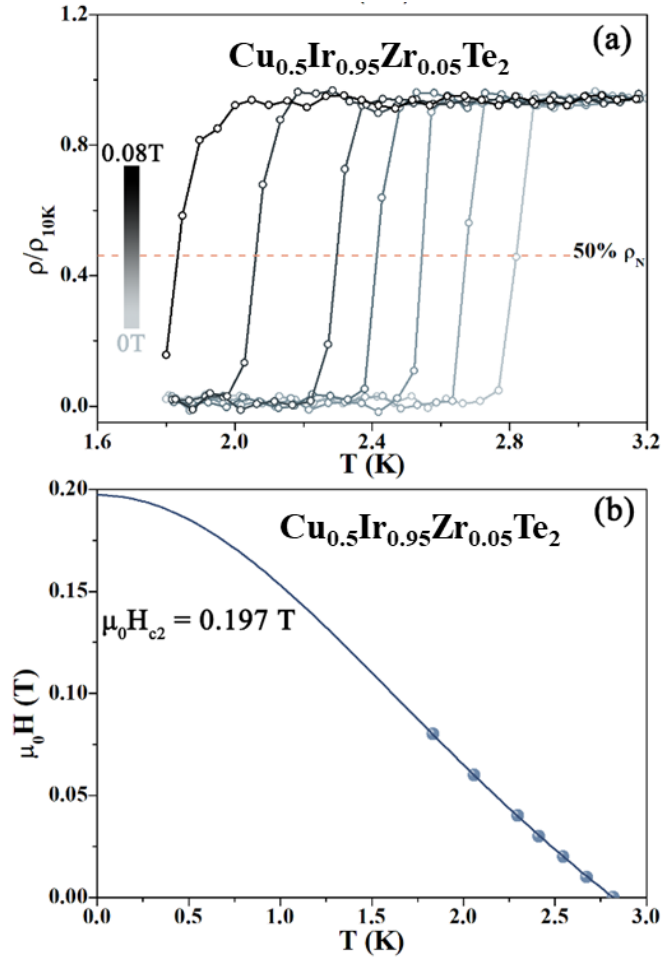

**Figure 6** (a) Low temperature resistivity of $Cu_{0.5}Ir_{0.95}Zr_{0.05}Te_2$ at various applied fields. (b) The $\mu_0H_{c1}(0)$ and $\mu_0H_{c2}(0)$ for $Cu_{0.5}Ir_{0.95}Zr_{0.05}Te_2$.

The upper critical field $\mu_0H_{c2}(0)$ is obtained through measuring the field dependence of $T_c$ in resistivity. The temperature-dependent electrical resistivity under various magnetic up to 0.08 T is illustrated in **Figure 6a** for the $Cu_{0.5}Ir_{0.95}Zr_{0.05}Te_2$ compound. The $T_c$ is gradually suppressed to a lower temperature as the field increases. For calculating the $\mu_0H_{c2}(0)$, the $T_c$ at 50 % criteria of the normal-state resistance have been used. It is evident from **Figure 6b** that $\mu_0H_{c2}(0)$ varies linearly with the temperature and can possibly be best fitted with Ginzburg-Landau equation given by $\mu_0 H_{c2}(T) = \mu_0 H_{c2}(0) * \frac{1-(T/T_c)^2}{1+(T/T_c)^2}$. Fitting with the 50 % $\rho_N$ criteria data can yield $\mu_0 H_{c2}(0) = 0.197$ T. Furthermore, we can acquire the $\mu_0 H_{c2}(0)$ value of 0.158 T for $Cu_{0.5}Ir_{0.95}Zr_{0.05}Te_2$ via using the WHH formula $\mu_0 H_{c2} = -0.693 T_c \frac{dH_{c2}}{dT_c}$ for the dirty limit SC[38]. The slopes, $dH_{c2}/dT$, are obtained from linear fitted $Cu_{0.5}Ir_{0.95}Zr_{0.05}Te_2$ sample for data 50 % criteria of $\rho_N$. Additionally, we fit the Pauli limiting field ($\mu_0 H^P(T)$) of $Cu_{0.5}Ir_{0.95}Zr_{0.05}Te_2$ by the formula $\mu_0 H^P(T) = 1.86 T_c$, which is shown in **Table 2**. We can calculated the Ginzburg Landau coherence length $\xi_{GL}(0)$ from the $\mu_0 H_{c2}(0)$ by the formulation $H_{c2} = \phi_0 / 2\pi \xi_{GL}^2$, where $\Phi_0$ represents the quantum flux (h/2e). We obtain $\xi_{GL}(0) = 45.6$ nm.

**Table 2.** Comparison of physical performance of $Cu_{0.5}IrTe_2$-based superconductors.

| Compound | $Cu_{0.5}Ir_{0.95}Zr_{0.05}Te_2$ | $Cu_{0.5}Ir_{0.9625}Ti_{0.0375}Te_2$[28] | $Cu_{0.5}Ir_{0.975}Ru_{0.025}Te_2$[27] | $Cu_{0.5}IrTe_2$[26] |
|---|---|---|---|---|
| $T_c$ (K) | 2.80 | 2.84 | 2.79 | 2.50 |
| $\gamma$ (mJ mol$^{-1}$ K$^{-2}$) | 12.57 | 14.13 | 12.26 | 12.05 |
| $\beta$ (mJ mol$^{-1}$ K$^{-4}$) | 1.86 | 2.72 | 1.87 | 1.97 |
| $\Theta_D$ (K) | 193.9(9) | 170.9(1) | 193.6(2) | 190.3(1) |
| $\Delta C/\gamma T_c$ | 1.45 | 1.34 | 1.51 | 1.50 |
| $\lambda_{ep}$ | 0.61 | 0.64 | 0.65 | 0.63 |
| $N(E_F)$ (states/eV f.u) | 3.31 | 3.67 | 3.15 | 3.10 |
| $\mu_0 H_{c2}(T)$ (50 % $\rho_N$ WHH theory) | 0.158 | 0.212 | 0.247 | 0.12 |
| $\mu_0 H^P(T)$ | 5.21 | 5.28 | 5.19 | 4.65 |
| $\mu_0 H_{c1}(T)$ | 0.039 | 0.095 | 0.098 | 0.028 |
| $\xi_{GL}(0)$ (nm) (50 % $\rho_N$ WHH theory) | 45.6 | 39.3 | 36.3 | 52.8 |

To further convince that SC is an intrinsic property of the optimal doping composition $Cu_{0.5}Ir_{0.95}Zr_{0.05}Te_2$, we also perform the temperature-dependent specific heat measurement under 0 and 0.08 T magnetic fields, as shown in **Figure 7**. The curve of $C_p/T$ vs $T^2$ exhibits significant jumps at low temperature under 0 T, signifying the bulk nature of SC and the high quality of the sample. Following equal-entropy construction, the estimated $T_c$ is 2.75 K, which is in agreement with resistance and magnetic susceptibility measurements. Besides, at temperatures between $T_c$ and 10 K, the specific heats are well described as T-linear electronic contribution and a sum of a $T^3$ phonon contribution. The curves can be fitted using the formula $C_p = \gamma T + \beta T^3$, where the $\gamma$ represents Sommerfeld's coefficient and $\beta$ is the phonon specific heat coefficient. The value of $\beta$ and $\gamma$ is calculated to be 1.56 mJ mol$^{-1}$ K$^{-4}$ and 12.57 mJ mol$^{-1}$ K$^{-2}$, respectively. And the normalized specific heat jump ($\Delta C/\gamma T_c$) value is further deduced to be 1.45, which is close to the weak-coupling limit value 1.43 of Bardeen-Cooper-Schrieffer (BCS) theory, proving bulk SC in our studied $Cu_{0.5}Ir_{0.95}Zr_{0.05}Te_2$ compound. On the other hand, the Debye temperatures ($\Theta_D$) further can be determined to be 185.7 K based on the expression $\Theta_D = (12\pi^4 nR/5\beta)^{1/3}$, where $n$ is the number of atoms per formula unit and $R$ represents ideal gas constant. Assuming $\mu^* = 0.13$, the electron-phonon coupling constant ($\lambda_{ep}$) is further calculated as 0.61 from the McMillan equation: $\lambda_{ep} = \frac{1.04 + \mu^* \ln\left(\frac{\Theta_D}{1.45 T_c}\right)}{(1-0.62\mu^*)\ln\left(\frac{\Theta_D}{1.45 T_c}\right) - 1.04}$[47]. Following the formula $N(E_F) = \frac{3}{\pi^2 k_B^2 (1+\lambda_{ep})}\gamma$, the $\gamma$ and $\lambda_{ep}$ values can give rise to the electron DOS at the Fermi level ($N(E_F)$) = 3.31.

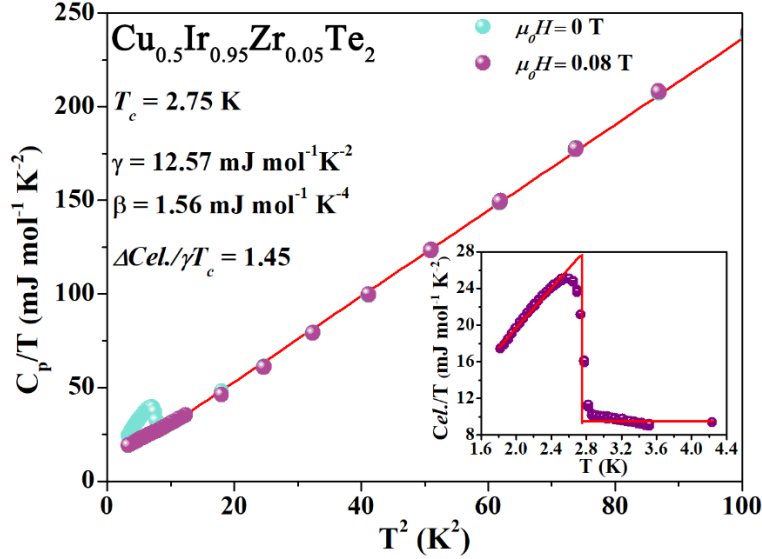

**Figure 7.** Detailed characterization of the superconducting transition in the highest $T_c$ compound $Cu_{0.5}Ir_{0.95}Zr_{0.05}Te_2$ through specific heat measurements under 0 and 0.08 T magnetic fields. Inset: the electronic contribution to the specific heat at the temperatures near the $T_c$ at zero magnetic field.

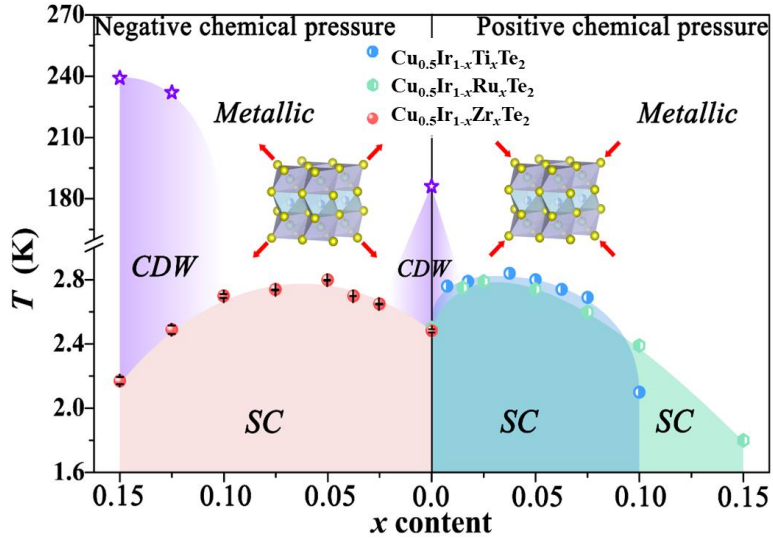

**Figure 8.** Electronic phase diagrams of $Cu_{0.5}Ir_{1-x}Zr_xTe_2$, $Cu_{0.5}Ir_{1-x}Ti_xTe_2$ (ref. 35), and $Cu_{0.5}Ir_{1-x}Ru_xTe_2$ (ref. 34) systems.

A schematic phase diagram of $Cu_{0.5}Ir_{1-x}Zr_xTe_2$ drawn based on the experimental results is shown in **Figure 8**. With increasing Zr concentration, both the SC and CDW phases were substantially affected. CDW order first quenches at $x = 0.025$, and $T_c$ gradually increases from 2.50 K to 2.80 K as $x \sim 0.05$. Once $x$ is above 0.05, the $T_c$ gradually decreases down to the largest Zr-doped content and a classic dome-like $T_c$ can be observed in the $Cu_{0.5}Ir_{1-x}Zr_xTe_2$ system. With increasing Zr content up to 0.125, interesting, the CDW phase reappears. It can be clearly seen that the CDW phase locates near both ends of the dome-shape SC phase, suggesting a complex interplay between the two phases, potentially supporting the Cooper pairing and CDW instabilities competing directly for the Fermi surface. The opening of the gap in the CDW state would significantly reduce the DOS at the Fermi surface, leading to the suppressing the SC in BCS scenario.

In TMDCs, the interaction between the SC and the CDW displays a complex phase diagram and even presents different phase diagrams from the same pristine compound through different tuning methods. Typically, competition, cooperation, or insensitivity between the CDW and the SC are considered to exist in TMDCs determined by various modulation methods[11-13, 20, 25]. Relatedly, it is worth discussing the distinct features of the phase diagrams of the $Cu_{0.5}IrTe_2$ system with other chemical doping. In the same Ir-site doping, the difference between Zr and Ru/Ti substitutions can be seen clearly in the phase diagram, combining Zr and Ru/Ti substitutions corresponding to negative and positive chemical pressure, respectively[34,35]. The CDW order in $Cu_{0.5}Ir_{1-x}Zr_xTe_2$ reappears for $x \geq 0.125$. This shows a clear contrast to the $Cu_{0.5}Ir_{1-x}Ru_xTe_2$ and $Cu_{0.5}Ir_{1-x}Ti_xTe_2$ system, where the CDW order is quickly suppressed and vanishes in all Ru/Ti-substituted compounds. Besides, in the Te-site doping, the CDW order in the $Cu_{0.5}IrTe_{2-x}I_x$ system also reemerges at $x \geq 0.45$[44]. It should be noted that I substitute Te causes the cell parameters to expand, i.e., it creates negative chemical pressure[41]. One possibility is that the high negative chemical pressure easier induces the disorder in the $Cu_{0.5}IrTe_2$ system. Negative chemical pressure causes an increase in the *c*-axis lattice parameter so that the coupling between the layers in $Cu_{0.5}IrTe_2$ could become weaker and the system becomes more two-dimensional, which could explain the reemergence and enhancement of the CDW instability in high Zr doping regions. Therefore, high negative chemical pressure seems to reemerge the CDW order in the $Cu_{0.5}IrTe_2$ system. Besides, the first suppression and then the reemergence of CDW order near the end of dome-shape SC phase diagram was also observed in the 2H-TaSe$_{2-x}$S$_x$ system, in which the disorder played an important role in the interplay of CDW and SC[23]. Finally, due to the highly complex disorder effects and impurity scattering in multiband SC as well as the coexistence of SC with a CDW order, the term dirty limit in the Cu$_{0.5}$IrTe$_2$ system may describe qualitatively quite different scenarios and its meaning for superconducting properties must be specified carefully.

**CONCLUSIONS**

In brief, we have designed and discussed the physical properties of Cu$_{0.5}$Ir$_{1-x}$Zr$_x$Te$_2$ ($0 \leq x \leq 0.15$) samples. XRD data indicates that the crystal lattice was significantly expanded by Zr substitution, signifying the negative chemical pressure effect. Upon Zr concentration, the CDW state was suppressed immediately and a classical dome-shaped $T_c(x)$ in doped regime can be observed. The optimal superconducting composition is Cu$_{0.5}$Ir$_{0.95}$Zr$_{0.05}$Te$_2$, which is identified as a type-II superconductor with $T_c \sim 2.8$ K. Once $x$ is above 0.05, the $T_c$ gradually decreases down to the largest Zr-doped content. When $x \geq 0.125$, the CDW order reemerges, and the $T_{CDW}$ increases, indicating strong competition between two phases. The increase in $T_{CDW}$ is directly correlated with crystallographic disorder and disorder induced scattering of the local CDW fluctuations.

**Notes**

The authors declare no competing financial interest

**ASSOCIATED CONTENT**

**Supporting Information**

The Supporting Information is available free of charge.

SEM, EDXS mappings and the element ratios of $Cu_{0.5}Ir_{1-x}Zr_xTe_2$ samples from EDXS results. Hall traces of three different $Cu_{0.5}Ir_{1-x}Zr_xTe_2$ ($x$ = 0.0325, 0.075, 0.1) samples and carrier density of the three different $Cu_{0.5}Ir_{1-x}Zr_xTe_2$ ($x$ = 0.0325, 0.075, 0.1) samples as a function of temperature.

**ACKNOWLEDGMENT**

This work was financially supported by the National Natural Science Foundation of China (No. 11922415), Guangdong Basic and Applied Basic Research Foundation (2019A1515011718), Key Research & Development Program of Guangdong Province, China (2019B110209003), the Pearl River Scholarship Program of Guangdong Province Universities and Colleges (20191001).

REFERENCES

1. Somayazulu, M.; Ahart, M.; Mishra, A. K.; Geballe, Z. M.; Baldini, M.; Meng, Y.; Struzhkin, V. V.; Hemley, R. J. Evidence for Superconductivity above 260 K in Lanthanum Superhydride at Megabar Pressures. *Phys. Rev. Lett.* **2019**, *122* (2), 027001.

2. Csontos, M.; Mihály, G.; Jankó, B.; Wojtowicz, T.; Liu, X.; Furdyna, J. K. Pressure-induced ferromagnetism in (In,Mn)Sb dilute magnetic semiconductor. *Nat. Mate.* **2005**, *4* (6), 447-449.

3. Aoyama, T.; Yamauchi, K.; Iyama, A.; Picozzi, S.; Shimizu, K.; Kimura, T. Giant spin-driven ferroelectric polarization in $TbMnO_3$ under high pressure. *Nat. Commun.* **2014**, *5* (1), 4927.

4. Wang, Y.; Zhang, L.; Wang, J.; Li, Q.; Wang, H.; Gu, L.; Chen, J.; Deng, J.; Lin, K.; Huang, L.; et al. Chemical-Pressure-Modulated $BaTiO_3$ Thin Films with Large Spontaneous Polarization and High Curie Temperature. *J. Am. Chem. Soc.* **2021**, *143* (17), 6491-6497.

5. Yamamoto, A.; Takeshita, N.; Terakura, C.; Tokura, Y. High pressure effects revisited for the cuprate superconductor family with highest critical temperature. *Nat. Commun.* **2015**, *6*, 8990.

6. Mukasa, K.; Matsuura, K.; Qiu, M.; Saito, M.; Sugimura, Y.; Ishida, K.; Otani, M.; Onishi, Y.; Mizukami, Y.; Hashimoto, K.; et al. High-pressure phase diagrams of FeSe$_{1-x}$Te$_x$: correlation between suppressed nematicity and enhanced superconductivity. *Nat. Commun.* **2021**, *12* (1), 381-381.

7. Chen, K. Y.; Wang, N. N.; Yin, Q. W.; Gu, Y. H.; Jiang, K.; Tu, Z. J.; Gong, C. S.; Uwatoko, Y.; Sun, J. P.; Lei, H. C.; et al. Double Superconducting Dome and Triple Enhancement of $T_c$ in the Kagome Superconductor CsV$_3$Sb$_5$ under High Pressure. *Phys. Rev. Lett.* **2021**, *126* (24), 247001.

8. Chi, Z.; Chen, X.; Yen, F.; Peng, F.; Zhou, Y.; Zhu, J.; Zhang, Y.; Liu, X.; Lin, C.; Chu, S.; et al. Superconductivity in Pristine 2H$_a$-MoS$_2$ at Ultrahigh Pressure. *Phys. Rev. Lett.* **2018,** *120* (3), 037002.

9. Chi, Z. H.; Zhao, X. M.; Zhang, H.; Goncharov, A. F.; Lobanov, S. S.; Kagayama, T.; Sakata, M.; Chen, X. J. Pressure-Induced Metallization of Molybdenum Disulfide. *Phys. Rev. Lett.* **2014**, *113* (3), 036802.

10. Pan, X. C.; Chen, X.; Liu, H.; Feng, Y.; Wei, Z. Pressure-driven dome-shaped superconductivity and electronic structural evolution in tungsten ditelluride. *Nat. Commun.* **2015**, *6*, 7805.

11. Sipos, B.; Kusmartseva, A.; Akrap, A.; Berger, H.; Tutis, E. J. N. M. From Mott state to superconductivity in-1T-TaS$_2$. *Nat. Mater.* **2008**, *7*, 960-965.
20

12. Joe, Y. I.; Chen, X. M.; Ghaemi, P.; Finkelstein, K. D.; De, L.; Gan, Y.; Lee, J.; Yuan, S.; Geck, J.; Macdougall, G. J. J. D.; et al. Emergence of charge density wave domain walls above the superconducting dome in TiSe$_2$. *Nat. Phys.* **2014**, *10*, 421-425.

13. Lin, T.; Wang, X.; Chen, X.; Liu, X.; Luo, X.; Li, X.; Jing, X.; Dong, Q.; Liu, B.; Liu, H.; et al. Retainable Superconductivity and Structural Transition in 1T-TaSe$_2$ Under High Pressure. *Inorg. Chem.* **2021,** *60*, 11385-11393.

14. Snider, E.; Dasenbrock-Gammon, N.; McBride, R.; Debessai, M.; Vindana, H.; Vencatasamy, K.; Lawler, K. V.; Salamat, A.; Dias, R. P. Room-temperature superconductivity in a carbonaceous sulfur hydride. *Nature* **2020**, *586* (7829), 373-377.

15. Bednorz, J. G.; Müller, K. A. Possible high $T_c$ superconductivity in the Ba-La-Cu-O system. *Z. Phys. B Condens. Matt.* **1986**, *64* (2), 189-193.

16. Lee, P. A.; Nagaosa, N.; Wen, X.-G. Doping a Mott insulator: Physics of high-temperature superconductivity. *Rev. Mod. Phys.* **2006**, *78* (1), 17-85.

17. Ren, Z. A.; Lu, W.; Yang, J.; Yi, W.; Shen, X. L.; Li, Z. C.; Che, G. G.; Dong, X. L.; Sun, L. L.; Zhou, F.; et al. Superconductivity at 55 K in Iron-Based F-Doped Layered Quaternary Compound SmO$_{1-x}$F$_x$FeAs. *Chinese. Phys. Lett.* **2008**, *25* (6), 2215-2216.

18. Hsu, F. C.; Luo, J. Y.; Yeh, K. W.; Chen, T. K.; Huang, T. W.; Wu, P. M.; Lee, Y. C.; Huang, Y. L.; Chu, Y. Y.; Yan, D. C.; et al. Superconductivity in the PbO-type structure *a*-FeSe. *Proc. Natl. Acad. Sci. USA* **2008**, *105* (38), 14262-14264.

19. Wang, X. C.; Liu, Q. Q.; Lv, Y. X.; Gao, W. B.; Yang, L. X.; Yu, R. C.; Li, F. Y.; Jin, C. Q. The superconductivity at 18 K in LiFeAs system. *Solid. State. Commun.* **2008**, *148* (11), 538-540.

20. Li, L. J.; Lu, W. J.; Zhu, X. D.; Ling, L. S.; Qu, Z.; Sun, Y. P. Fe-doping–induced superconductivity in the charge-density-wave system 1T-TaS$_2$. *Europhys. Lett.* **2012**, *97*, 6.

21. Bhoi, D.; Khim, S.; Nam, W.; Lee, B. S.; Kim, C.; Jeon, B. G.; Min, B. H.; Park, S.; Kim, K. H. Interplay of charge density wave and multiband superconductivity in 2H-Pd$_x$TaSe$_2$. *Sci. Rep.* **2016**, *6*, 24068.

22. Li, L.; Deng, X.; Wang, Z.; Liu, Y.; Abeykoon, M.; Dooryhee, E.; Tomic, A.; Huang, Y.; Warren, J. B.; Bozin, E. S.; et al. Superconducting order from disorder in 2H-TaSe$_{2-x}$S$_x$. *npj Quantum Mater.* **2017**, *2* (1), 11.

23. Luo, H. X.; Xie, W. W.; Tao, J.; Inoue, H.; Gyenis, A.; Krizan, J. W.; Yazdani, A.; Zhu, Y. M.; Cava, R. J. Polytypism, polymorphism, and superconductivity in TaSe$_{2-x}$Te$_x$. *Proc. Natl. Acad. Sci. USA* **2015**, *112* (11), E1174-80.

24. Cho, K.; Kończykowski, M.; Teknowijoyo, S.; Tanatar, M. A.; Guss, J.; Gartin, P. B.; Wilde, J. M.; Kreyssig, A.; McQueeney, R. J.; Goldman, A. I.; et al. Using controlled disorder to probe the interplay between charge order and superconductivity in NbSe$_2$. *Nat. Commun.* **2018,** *9* (1), 2796-2796.

25. Morosan, E.; Zandbergen, H. W.; Dennis, B. S.; Bos, J.; Onose, Y.; Klimczuk, T.; Ramirez, A. P.; Ong, N. P.; Cava, R. J. Superconductivity in Cu$_x$TiSe$_2$. *Nat. Phys.* **2006**, *2* (8), 544-550.

26. Matsumoto, N.; Taniguchi, K.; Endoh, R.; Takano, H.; Nagata, S. Resistance and Susceptibility anomalies in IrTe$_2$ and CuIr$_2$Te$_4$. *J. Low Temp. Phys*. **1999**, *117*, 1129.

27. Cao, H. B.; Chakoumakos, B. C.; Chen, X.; Yan, J.-Q.; McGuire, M. A.; Yang, H.; Custelcean, R.; Zhou, H. D.; Singh, D. J.; Mandrus, D. G. Origin of the phase transition in IrTe$_2$: structural modulation and local bonding instability. *Phys. Rev. B* **2013**, *88*, 115122.

28. Yan, J.-Q.; Saparov, B.; Sefat, A. S.; Yang, H.; Cao, H. B.; Zhou, H. D.; Sales, B. C.; Mandrus, D. G. Absence of structural transition in $M_{0.5}$IrTe$_2$ ($M$ = Mn, Fe, Co, Ni). *Phys. Rev. B* **2013**, *88*, 134502.

29. Yang, J. J.; Choi, Y. J.; Oh, Y. S.; Hogan, A.; Horibe, Y.; Kim, K.; Min, B. I.; Cheong, S.-W. Charge-oribital density wave and superconductivity in the strong spin-orbit coupled IrTe$_2$ : Pd. *Phys. Rev. Lett*. **2012**, *108*, 116402.

30. Yu, R.; Banerjee, S.; Lei, H.; Abeykoon, M.; Petrovic, C.; Guguchia, Z.; Bozin, E. S. Phase separation at the dimer-superconductor transition in Ir$_{1-x}$Rh$_x$Te$_2$. *Phys. Rev. B* **2018**, *98*, 134506.

31. Pyon, S.; Kudo, K.; Nohara, M. Emergency of superconductivity near the structural phase boundary in Pt-doped IrTe$_2$ single crystals. *Physica C* **2013**, *494*(11), 80-84.

32. Mamitani, M.; Bahramy, M. S.; Arita, R.; Seki, S.; Arima, T.; Tokura, Y.; Ishiwata, S. Superconductivity in Cu$_x$IrTe$_2$ driven by interlayer hybridization. *Phys. Rev. B* **2013**, 87, 180501(R).

33. Yan, D.; Zeng, Y.; Wang, G.; Liu, Y.; Yin, J.; Chang, T.-R.; Lin, H.; Wang, M.; Ma, J.; Jia, S.; et al. CuIr$_2$Te$_4$: A Quasi-Two-Dimensional Ternary Telluride Chalcogenide Superconductor. *arXiv:1908.05438*.

34. Yan, D.; Zeng, L. Y.; Lin, Y. S.; Yin, J. J.; He, Y. Y.; Zhang, X.; Huang, M. L.; Shen, B.; Wang, M.; Wang, Y. H.; et al. Superconductivity in Ru-doped CuIr$_2$Te$_4$ telluride chalcogenide. *Phys. Rev. B* **2019**, *100*, 174504.

35. Zeng, L. Y.; Yan, D.; He, Y. Y.; Boubeche, M.; Huang, Y. H.; Wang, X. P.; Luo, H. X. Effect of Ti substitution on the superconductivity of CuIr$_2$Te$_4$ telluride chalcogenide. *J. Alloy. Compd.* **2021**, *885*, 160981.

36. Bhatia, M.; Sumption, M.; Collings, E.; Dregia, S. Increases in the Irreversibility Field and the Upper Critical Field of Bulk MgB$_2$ by ZrB$_2$ Addition. *Appl. Phys. Lett.* **2005**, *87*, 042505.

37. Ban, E.; Ikebe, Y.; Matsuoka, Y.; Nishijima, G.; Watanabe, K. Transport Critical Current of Filamentary Zr-Doped Gd-Ba-Cu-O Superconductors in High Magnetic Fields. *Applied Superconductivity, IEEE Transactions on* **2008**, *18*, 1200-1203.

38. Xu, C.; Hu, A.; Ichihara, M.; Sakai, N.; Izumi, M.; Hirabayashi, I. Enhanced flux pinning in GdBaCuO bulk superconductors by Zr dopants. *Physica. C* **2007**, *463-465*, 367-370.

39. Xia, J. A.; Strickland, N. M.; Talantsev, E. F.; Long, N. J. Formation of Nanoparticles in Zr and Dy Doped YBCO MOD Superconducting Films. *Mater. Sci. Forum.* **2011**, *700*, 15-18.

40. Cortés, S.; Seaman, O. D. L. P. Electron- and hole-doping on ScH$_2$ and YH$_2$: Eects on superconductivity without applied pressure. J. *Phys-Condens. Mat*. **2021**, *33*, 425401.

41. Clementi, E.; Raimondi, D. Atomic Screening Constants from SCF Functions. *J. Chem. Phys.* **1963**, *38*, 2686-2689.

42. Chemical Bonding and Lattice Energy. In *Inorganic Structural Chemistry*, **2006**, 39-44.

43. Zhang, P.; Zhai, H. F.; Tang, Z. J.; Li, L.; Li, Y. K.; Chen, Q.; Chen, J. ; Wang, Z.; Feng, C. M.; Cao, G. H.; et al. Superconductivity enhanced by Se doping in Eu$_3$Bi$_2$(S,Se)$_4$F$_4$. *Europhys. Lett.* **2015**, *111*(2), 27002.

44. Boubeche, M.; Yu, J.; Li, C. S.; Wang, H. C.; Zeng, L. Y.; He, Y. Y.; Wang, X. P.; Su, W. Z.; Wang, M.; Yao, D. X.; et al. Superconductivity and Charge Density Wave in Iodine-Doped CuIr$_2$Te$_4$. *Chinese. Phys. Lett*. **2021**, *38* (3), 037401.

45. Ang, R.; Miyata, E.; Ieki, K.; Nakayama, K.; Sato, T.; Liu, Y.; Lu, W. J.; Sun, Y. P.; Takahashi, T. Superconductivity and bandwidth-controlled Mott metal-insulator transition in 1T-TaS$_{2-x}$Se$_x$ *Phys. Rev. B* **2013**, *88*, 115145

46. Wang, H. T.; Li, L. J.; Ye, D. S.; Cheng, X. H.; Xu, Z. A. Effect of Te doping on superconductivity and charge-density wave in dichalcogenides 2H-NbSe$_{2-x}$Te$_x$ ($x$ = 0, 0.1, 0.2). *Chinese. Phys*. **2007,** *16* (8), 4.

47. Bardeen, J.; Cooper, L. N.; Schrieffer, J. R. Theory of Superconductivity. *Phys. Rev.* **1957**, *108* (4), 717-726.

**TOC Graphic**

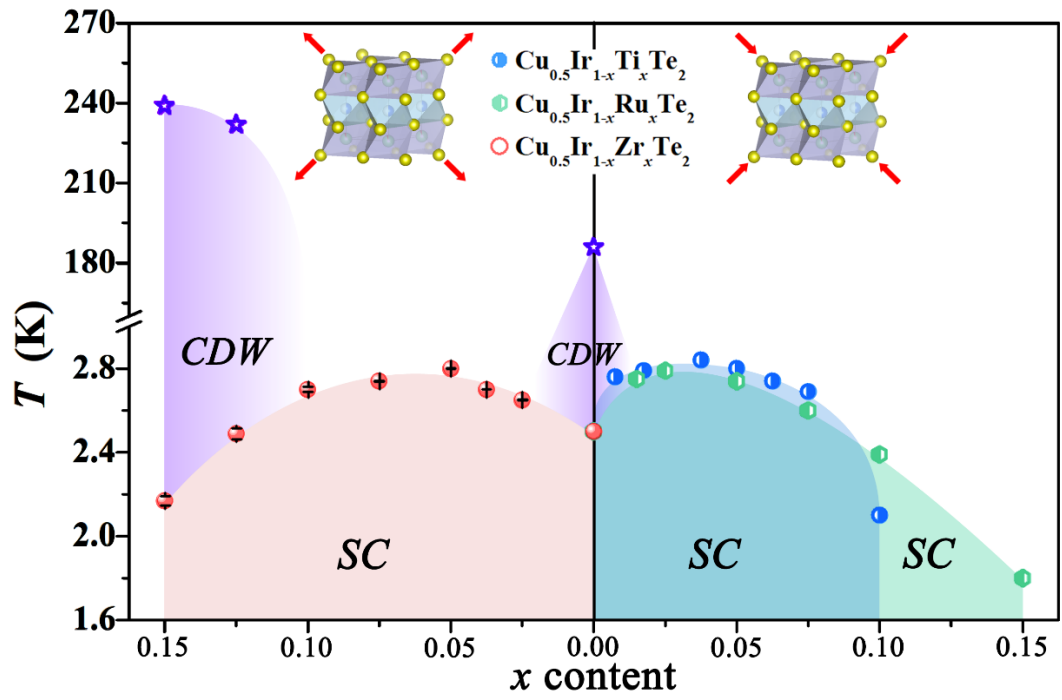

# Supporting information

# Negative Chemical Pressure Effect on Superconductivity and Charge Density Wave of $Cu_{0.5}Ir_{1-x}Zr_xTe_2$

*Lingyong Zeng[a], Yi Ji[b], Dongpeng Yu[a], Shu Guo[c,d], Yiyi He[a], Kuan Li[a], Yanhao Huang[a], Chao Zhang[a], Peifeng Yu[a], Shaojuan Luo[e], Huichao Wang[b], Huixia Luo[a*]*

[a]School of Materials Science and Engineering, State Key Laboratory of Optoelectronic Materials and Technologies, Key Lab of Polymer Composite & Functional Materials, Guangzhou Key Laboratory of Flexible Electronic Materials and Wearable Devices, Sun Yat-Sen University, No. 135, Xingang Xi Road, Guangzhou, 510275, P. R. China

[b]School of Physics, Sun Yat-Sen University, No. 135, Xingang Xi Road, Guangzhou, 510275, P. R. China

[c] Shenzhen Institute for Quantum Science and Engineering, Southern University of Science and Technology, Shenzhen 518055, China

[d]International Quantum Academy, Shenzhen 518048, China.

[e]School of Chemical Engineering and Light Industry, Guangdong University of Technology, Guangzhou, 510006, P. R. China

*Corresponding author/authors complete details (Telephone; E-mail:) (+86)-2039386124; E-mail address: luohx7@mail.sysu.edu.cn;

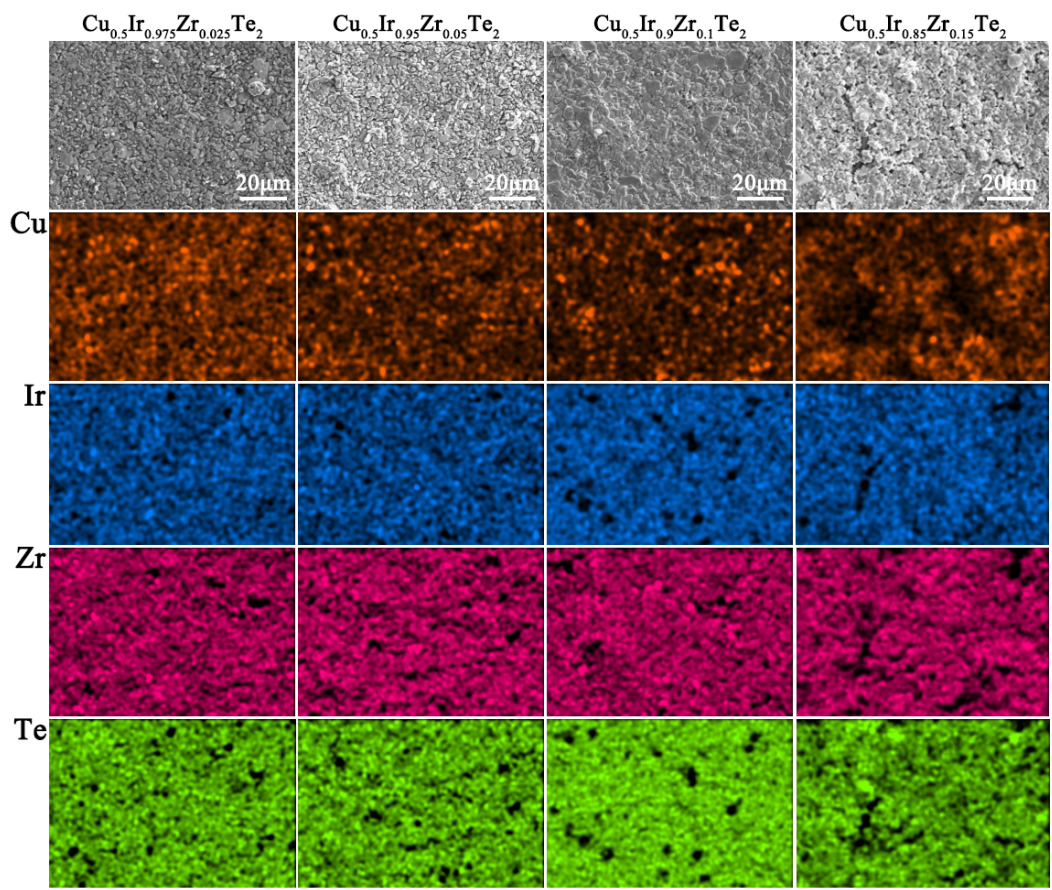

**Figure S1.** SEM and EDXS mappings of $Cu_{0.5}Ir_{1-x}Zr_xTe_2$ samples.

**Table S1**. The element ratios of $Cu_{0.5}Ir_{1-x}Zr_xTe_2$ from EDXS results.

| Sample \ Element ratio | Cu | Ir | Zr | Te |
|---|---|---|---|---|
| $Cu_{0.5}IrTe_2$ [1] | 0.485 | 0.980 | 0 | 1.965 |
| $Cu_{0.5}Ir_{0.975}Zr_{0.025}Te_2$ | 0.480 | 0.971 | 0.027 | 2.030 |
| $Cu_{0.5}Ir_{0.95}Zr_{0.05}Te_2$ | 0.483 | 0.932 | 0.056 | 2.035 |
| $Cu_{0.5}Ir_{0.9}Zr_{0.1}Te_2$ | 0.490 | 0.875 | 0.108 | 2.017 |
| $Cu_{0.5}Ir_{0.85}Zr_{0.15}Te_2$ | 0.486 | 0.838 | 0.161 | 2.026 |

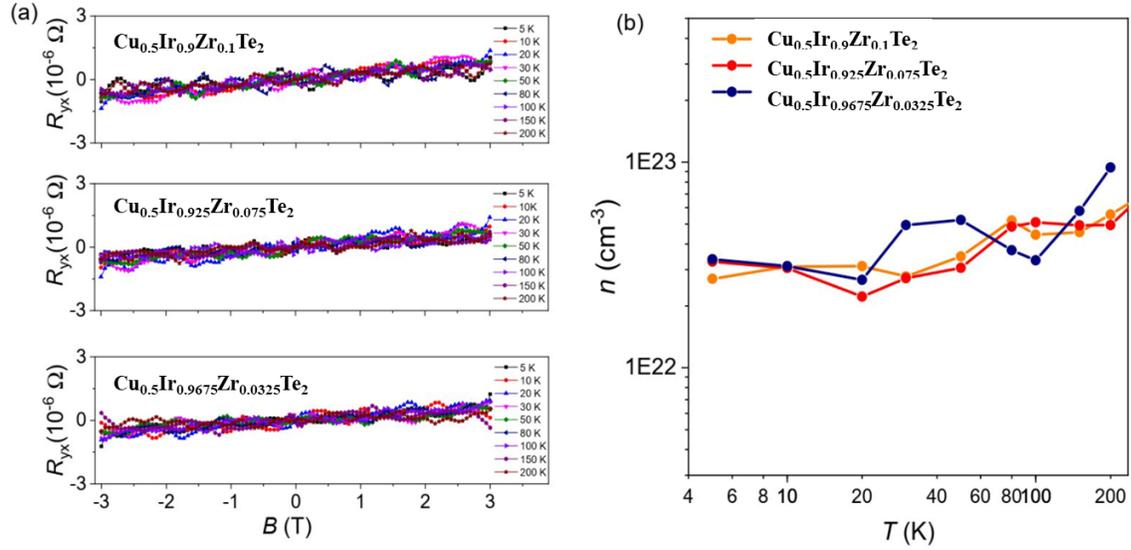

**Figure S2** (a) Hall traces of three different $Cu_{0.5}Ir_{1-x}Zr_xTe_2$ ($x$ = 0.0325, 0.075, 0.1) samples. (b) Carrier density of the three different $Cu_{0.5}Ir_{1-x}Zr_xTe_2$ ($x$ = 0.0325, 0.075, 0.1) samples as a function of temperature.